\documentclass{article}
\usepackage{spconf,amsmath,graphicx,hyperref,bm}
\usepackage{amsfonts,comment,pifont,booktabs}
\usepackage[table]{xcolor}
\usepackage{placeins}

\definecolor{lightgrey}{rgb}{0.925, 0.925, 0.925}

\renewcommand{\vec}[1]{\bm{#1}}

\title{SYMBIOTIC ARCHITECTURE FOR POST-HOC AUDIO EXTENSION OF FROZEN LANGUAGE MODELS}

\name{Yotaro Kubo, Qi Sun, Yujin Tang}
\address{Sakana AI, Tokyo, Japan.\\
\texttt{\{yotarokubo,qisun,yujin\}@sakana.ai}
}

\begin{document}
\ninept
\maketitle

\begin{abstract}
This paper proposes an architecture for equipping large language models (LLMs) with audio-understanding capabilities without fine-tuning their weights. The proposed symbiotic architecture employs an injector module that writes audio-conditioned vectors directly into the target LLM's short-term memory, i.e., the key-value (KV) cache, enabling the LLM to behave as an audio language model (ALM). The architectural advantages are twofold. First, it improves the scalability of ALMs: because the proposed method bypasses the LLM during audio injection, the injection cost is governed by the injector width rather than the backbone width, and can therefore scale more slowly than the cost of full-backbone prefilling. Second, since the training scheme does not update the LLM weights, the original capabilities of the LLM are preserved without the risk of degradation from fine-tuning. The effectiveness of the proposed method is evaluated on both audio-understanding tasks (automatic speech recognition, audio question answering, and acoustic scene classification) and text-only tasks. We confirm that, while activating fewer parameters during audio prefilling, our architecture outperforms the conventional method with a frozen LLM and approaches the performance of a fine-tuned ALM, all while preserving the backbone LLM's original text-only task performance by construction.
\end{abstract}
\begin{keywords}
Audio language model, context optimization, prefilling, catastrophic forgetting, short-term memory.
\end{keywords}

\begin{figure*}[!tb]
  \begin{center}
    \includegraphics[width=0.9\linewidth]{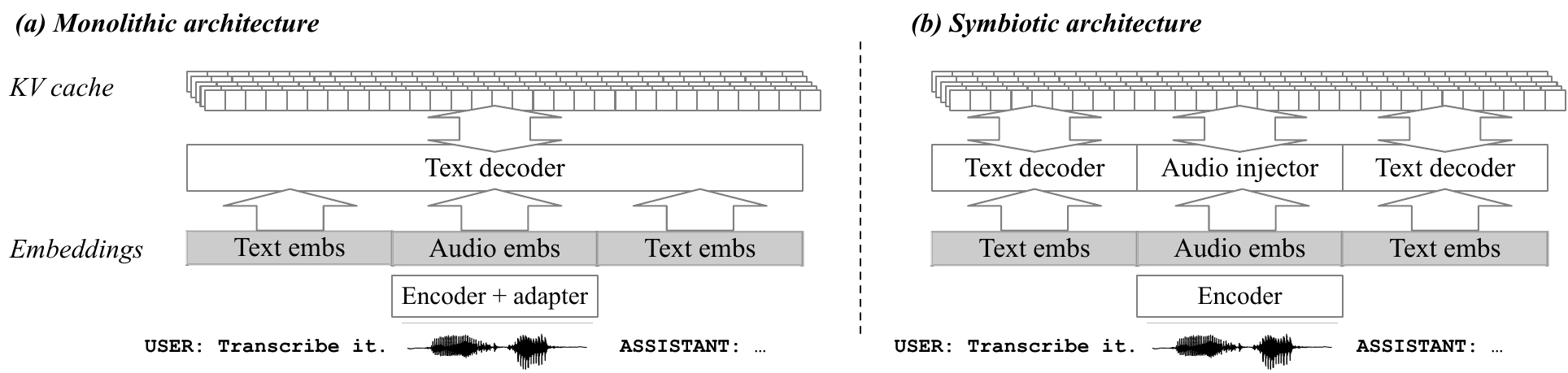}
    \caption{
    The conventional ``monolithic'' architecture and the proposed ``symbiotic'' architecture, which can accommodate both a large decoder and a fast, small injector.
    \label{fig:block_diagram}
    }
    \vspace{-6pt}
  \end{center}
\end{figure*}

\section{Introduction} \label{sect:introduction}

Audio language models (ALMs)~\cite{latif2023sparks} are a promising approach to enabling machines to understand general audio signals.
By leveraging the strong language-understanding capabilities of large language models (LLMs), an ALM can serve as a general-purpose solution for a wide range of audio-understanding tasks, including automatic speech recognition (ASR), audio question answering (AQA), and acoustic scene classification (ASC)~\cite{barchiesi2015acoustic}.

A typical ALM consists of three components: an input encoder, an adapter, and a backbone LLM.
The input encoder converts raw audio into a hidden representation.
The adapter, typically a very thin neural network, transforms the hidden representation into embedding vectors that are compatible with the backbone LLM.
Finally, 
the LLM autoregressively models interleaved sequences of text and audio embedding vectors.
To inherit the strong language-understanding capabilities of LLMs, a model pretrained on vast amounts of text data is adopted as a seed model and then fine-tuned to handle audio embedding vectors that were unseen during pretraining.
Although this monolithic architecture is advantageous in its simplicity, we identify two key drawbacks: limited prefilling scalability and catastrophic forgetting.

\textbf{(a) Limited prefilling scalability:}
In a typical ALM, audio data is provided as part of the input sequence and is passed to the response generation phase via a short-term memory implemented in the form of a key-value (KV) cache.
The KV cache is read later when the LLM generates the response text.
The conversion process from embedding vectors to the KV cache is called ``prefilling.''
Since this process is performed solely by the LLM in the conventional monolithic architecture, the computational cost of prefilling is proportional to the size of the backbone LLM.
Because audio token sequences tend to be long in typical audio-understanding tasks, this prefilling cost is not negligible.
More importantly, given the current trend toward ever-larger backbone LLMs, most of this computation is unrelated to audio processing; prefilling should therefore be decoupled from the backbone so that its cost no longer scales with the backbone size.

\textbf{(b) Catastrophic forgetting:}
The typical training process of an ALM involves supervised fine-tuning with audio and instruction data.
Adding a supervised training phase often degrades capabilities acquired during previous training phases.
This phenomenon is known as ``catastrophic forgetting.''
To avoid this, it is common practice to employ a carefully crafted mixture of datasets that includes tasks from earlier training phases.
However, using such a multimodal data mixture can make this post-training process more complicated and computationally expensive.

In this paper, the above two drawbacks are addressed by introducing a novel architecture.
In the proposed architecture, we introduce a more expressive adapter, called the injector, that directly writes the input representations into the short-term memory, specifically the KV cache of the LLM.
Prefilling scalability is then improved by delegating prefilling to this injector, which can be a narrower model with fewer parameters than the LLM.
Motivated by the ability of contextual representations to steer frozen language models~\cite{pryzant2023automatic}, we formulate audio adaptation as the generation of an input-dependent, layer-wise KV cache rather than an update to the backbone weights.
This allows the injector to optimize the KV cache instead of updating the LLM weights, thereby preventing degradation of the frozen backbone caused by audio fine-tuning.

\section{Related Work}

The proposed architecture can be viewed as a Mixture-of-Experts (MoE) model \cite{lepikhin2021gshard} with routing predetermined by the input modality.
Like MoE models, this architecture offers the advantage of activating only a subset of parameters at any given time.
MoST \cite{lou2026most} extends the MoE architecture by selecting expert modules based on the modality (audio or text) of each token.
Our approach shares the core idea of using different modules depending on the input modality.
However, because MoST is built on the MoE framework of the backbone LLM, the prefilling cost is still dominated by the backbone LLM.

The prefilling cost also depends on the frame rate of the input audio embedding vectors.
To reduce the number of input frames, several subsampling techniques have been proposed.
SSR-Connector~\cite{tan2025ssr} integrates an alignment module to perform adaptive subsampling based on segments aligned to text tokens.
SALMONN~\cite{tang2024salmonn} integrates Q-Former~\cite{li2023blip} to obtain a fixed-length representation for each segment, which can also be viewed as adaptive subsampling.
The proposed method may further improve prefilling efficiency when used in combination with these advanced subsampling techniques.

Whereas we integrate an expressive injector, several efforts instead aim to make the audio encoder and adapter more lightweight.
SLM~\cite{wang2023slm} demonstrated that optimizing only a lightweight adapter suffices to train a speech LLM.
Because the LLM parameters remain unchanged while the model is adapted to audio tasks, this method inherently mitigates catastrophic forgetting.
Taking this direction to an extreme, Gemma~4~\cite{team2026gemma4} adopts an encoder-free approach that maps audio signals directly to LLM inputs with only an affine transformation.
In this paper, we compare our method with SLM and show SLM's limitations on generic (non-speech) audio tasks.

For post-hoc multimodal extension, LLaMA-Adapter~\cite{zhang2023llama} optimizes a fixed-length KV prefix augmented by the output of an image encoder.
In terms of how it controls the backbone LLM, this approach is similar to the proposed method.
Although LLaMA-Adapter can also keep the LLM parameters completely unchanged, it requires modification to the computation graph and is therefore not frozen from a computational standpoint, unlike the proposed method.
We aim for a post-hoc extension that can reuse even the existing software infrastructure developed for the backbone LLM.

\section{Symbiotic Architecture}

Fig.~\ref{fig:block_diagram} illustrates the differences between the conventional monolithic architecture and the proposed architecture for ALMs.
In the monolithic architecture, input audio signals are first converted into audio embeddings.
This step often involves a thin adapter layer to adjust the feature dimensionality and a subsampling module to reduce the number of embeddings.
Even with the subsampling module applied, the process of audio prefilling remains computationally expensive because the backbone LLM itself must transform audio embeddings into the KV cache.

In contrast, our architecture uses the injector to transform audio embeddings into the KV cache.
This decoupling lets the injector be much narrower than the backbone LLM, so the cost of audio injection does not scale with the LLM width, enabling the use of arbitrarily large LLMs without a proportional increase in injection cost.

Additionally, this architecture leverages the flexibility of context optimization by integrating the injector as a context controller.
Unlike the monolithic architecture, the proposed architecture can introduce arbitrary vectors into the KV cache while keeping the LLM weights untouched.
This adaptation scheme can effectively transform an LLM trained solely on text datasets into an ALM without fine-tuning the LLM, addressing the two aforementioned problems.

\section{Injector Module}

\begin{figure*}[!tb]
  \begin{center}
    \centerline{\includegraphics[width=\linewidth]{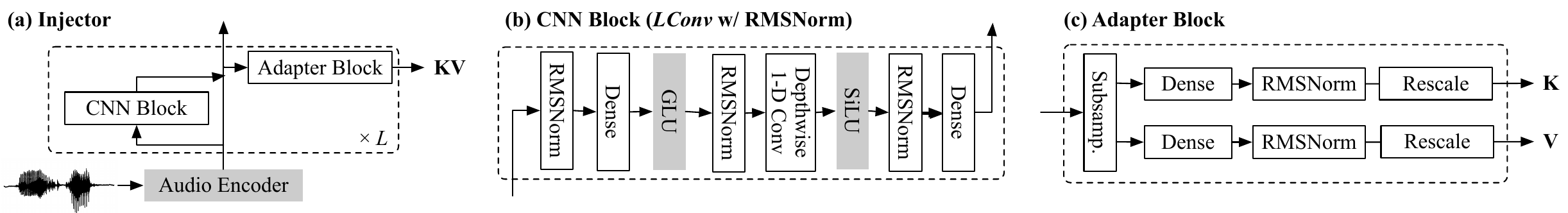}}
    \caption{
    Block diagrams of (a) the injector model, (b) the CNN block, and (c) the adapter block.
    }
    \label{fig:modules}
    \vspace{-6pt}
  \end{center}
\end{figure*}

Fig.~\ref{fig:modules} shows the block diagram of our injector module. 
Because the injector must produce a KV cache for every LLM layer,
it is designed to have the same number of layers as the backbone LLM.

\subsection{CNN Block}

The injector module is a stacked convolutional neural network (CNN) based on ``LConv'' modules, which play a crucial role in the Conformer architecture \cite{gulati2020conformer}.
An LConv module consists of a depthwise convolution flanked by two pointwise dense layers.
Note that we replace normalization layers in Conformer's LConv with RMSNorm to improve stability during small-batch training.
In the experiments, we feed 2,048-dimensional vectors into the depthwise 1-D convolution with a kernel size of 15.

In this paper, we employ a CNN-based architecture rather than a vanilla Transformer because we hypothesize that local context is more important than global context in audio prefilling.
Because the backbone LLM is a Transformer capable of integrating global context, the injector is designed as a complementary module that focuses on local context.
In preliminary experiments, we found that optimization was unstable when self-attention modules were used instead of ``LConv''.

\subsection{Adapter Block}

In the adapter block, the outputs of CNN layers are subsampled and mapped into the key and value vector spaces of the backbone LLM.

The adapter block first applies max pooling to subsample the sequence with a fixed stride of 8.
Then, the key and value vectors are extracted using two affine transformations (``Dense'' modules).
As in Qwen3, RMSNorm layers shared among multiple attention heads are applied to key vectors.
To stabilize training, RMSNorm modules with the same parametrization are also applied to the value vectors.

\subsection{KV Scale Matching}

To facilitate optimization, especially during the early phase of training, the key vectors generated by the injector must attract high attention scores; otherwise, too little gradient reaches the injector, and optimization plateaus.
In our experiments, we use Qwen3 \cite{yang2025qwen3} as the backbone LLM. Therefore, the distribution of the KV cache produced by the injector module should match that of Qwen3's KV cache.
The details in this section are specific to the Qwen3 backbone; however, the scale-matching strategy can also be applied to other backbones.

The matching is done by adjusting the initialization of the scale parameters of RMSNorm and introducing additional head-wise scale parameters.
Let $\vec{k}^{(\ell)}_{h, t}$ and $\vec{v}^{(\ell)}_{h, t}$ be the output key and value vectors from the injector for the $h$-th head in the $\ell$-th layer at time step $t$.
To match their distributions to those of the LLM's internal key and value vectors, RMSNorm is applied as follows:

\begin{equation}
\begin{aligned}
\vec{k}^{(\ell)}_{h, t} :=& \vec{\alpha}^{(\ell)}_h \odot \mathrm{RMSNorm}({\vec{k}}^{(\ell)}_{h, t}; \vec{s}^{(\ell)}),  \\
\vec{v}^{(\ell)}_{h, t} :=& \vec{\beta}^{(\ell)}_h \odot \mathrm{RMSNorm}({\vec{v}}^{(\ell)}_{h, t}; \vec{r}^{(\ell)}),
\end{aligned}
\end{equation}
where $\vec{s}^{(\ell)}$ and $\vec{r}^{(\ell)}$ are the scale parameters of RMSNorm modules.

The objective of scale matching is to adjust the magnitudes of $\vec{k}^{(\ell)}_{h, t}$ and $\vec{v}^{(\ell)}_{h, t}$ to match those in the backbone LLM.
Because Qwen3 also normalizes its key vectors using RMSNorm with scale parameters shared across attention heads, scale matching for $\vec{k}^{(\ell)}_{h,t}$ can be performed by initializing $\vec{s}^{(\ell)}$ with the corresponding Qwen3 RMSNorm scale parameters.
For $\vec{r}^{(\ell)}$, we empirically obtained the initial values for those scales by feeding a few calibration sentences into Qwen3 and computing the standard deviation of the elements of the value vectors.
The calibration sentences were designed to emulate instruction-response text pairs for typical audio tasks.

Because we share RMSNorm's scale parameters across attention heads, the RMSNorm outputs have limited per-head flexibility.
To compensate for this, we introduce additional scale parameters, initialized to 1, to rescale the RMSNorm output.

\subsection{Noisy RoPE Training}

RoPE \cite{su2024roformer} is known to make models overfit to the sequence lengths seen during training.
If the backbone LLM uses RoPE, the injector must also apply RoPE and thus inherit this fragility.
DroPE \cite{gelberg2026extending} alleviates this problem by removing RoPE during LLM training, but it is inapplicable here because our method does not fine-tune the backbone LLM.
To improve robustness to inputs with unseen lengths, we introduce a simple randomized training scheme for RoPE.

Let $x$ and $y$ denote the $2i$-th and $(2i + 1)$-th elements of $\vec{k}^{(\ell)}_{h, t}$.
RoPE applies the following transformation:

\begin{equation}
\begin{pmatrix}
x \\
y
\end{pmatrix}
:=
\begin{pmatrix}
\cos \left( t' \theta_i \right) & -\sin \left( t' \theta_i \right) \\
\sin \left( t' \theta_i \right) & \cos \left( t' \theta_i \right)
\end{pmatrix}
\begin{pmatrix}
x \\ y
\end{pmatrix},
\end{equation}
where $t' = t_0 + \tau + \kappa t$, and $\theta_i$ is the angular velocity for the $i$-th dimension pair as defined in~\cite{su2024roformer}.
Here, $t_0$ is the position offset for each instance,
$\tau$ is a random offset variable drawn from the uniform integer distribution over $[0, \tau_{\mathrm{MAX}}]$, and $\kappa$ is a random time-scale factor drawn from a uniform distribution over $[\kappa_{\mathrm{MIN}}, \kappa_{\mathrm{MAX}})$.

The conventional RoPE is a special case of this noisy RoPE, obtained by setting $\tau = 0$ and $\kappa = 1$.
The randomization introduced here follows the position-perturbation technique of~\cite{yu2025comrope} and improves robustness to input sequences of unseen lengths.
In the experimental section, we used $\tau_{\mathrm{MAX}} = 256$, $\kappa_{\mathrm{MIN}} = 0.75$, and $\kappa_{\mathrm{MAX}} = 1.5$.

\begin{table*}[!t]
\vspace{-6pt}
\caption{Error rates (\%) on audio and speech tasks. ``Audio'' and ``Text'' denote the number of parameters involved during audio prefilling and text prefilling, respectively.} 
\label{table:mler}
\centering
\begin{tabular}{lrrrrrrrrr}
\toprule
& 
\multicolumn{3}{c}{\bf \#Params} &
\multicolumn{4}{c}{\bf LibriSpeech} &
{\bf Clotho-AQA} &
{\bf CochlScene} 
\\
\cmidrule(lr){2-4}
\cmidrule(lr){5-8}
{\bf Architecture} &
Total & Audio & Text &
dev-clean & dev-other & test-clean & test-other 
\\
\midrule 
Encoder-only
& \textit{1108M} 
& \textit{1108M}
& \textit{752M}
& 2.22
& 4.07
& 2.33
& 4.25
& 62.6
& 52.9
\\
Symbiotic
& \textit{1303M} 
& \textit{551M}
& \textit{752M}
& {\bf 1.91}
& 3.73
& {\bf 2.07}
& {\bf 3.91}
& 47.1
& 41.7
\\
\midrule
Monolithic 
& \textit{1108M} 
& \textit{1108M}
& \textit{752M}
& 2.16
& {\bf 3.68}
& 2.26
& 4.01
& {\bf 45.4}
& {\bf 32.7}
\\
\bottomrule
\end{tabular}
\end{table*}

\section{Experiments}

\subsection{Model Configuration}

The pretrained WavLM (\texttt{wavlm-large}) model~\cite{chen2022wavlm} was adopted as our base audio encoder.
During training, LoRA was applied to this encoder module with the parameters $r=64, \alpha=64$.
As mentioned in the previous section, Qwen3~\cite{yang2025qwen3} was adopted as our LLM component.
Specifically, \texttt{Qwen3-0.6B} was chosen to evaluate the effectiveness of the proposed approach for compact ALMs.

For comparison, we built a model based on the SLM~\cite{wang2023slm} architecture using the same backbone LLM (\texttt{Qwen3-0.6B}) and the same audio encoder (\texttt{wavlm-large}).
We refer to this baseline as ``Encoder-only''.
We also evaluated ``Monolithic,'' a reference that adapts the backbone LLM with LoRA ($r=16,\alpha=16$).
In both conventional models, the connector stacks eight consecutive 1,024-dimensional audio representations and maps the resulting 8,192-dimensional vector into the LLM's 1,024-dimensional embedding space using an affine transformation.
This gives the same subsampling rate as the symbiotic architecture.

\subsection{Training Configuration}

\begin{table}[!t]
\vspace{-6pt}
\caption{Training datasets.}
\label{table:trainsets}
\centering
\begin{tabular}{lrcl}
\toprule
{\bf Dataset} & {\bf \#hrs} & {\bf Mix weight} & {\bf Task} \\
\midrule
LibriSpeech \cite{panayotov2015librispeech} & 960 & 90\% & ASR \\
CompA-R \cite{ghosh2024gama} & 159 & 2.5\% & AQA \\
DCASE2025 Task5 \cite{yang2025dcase} & 7.4 & 2.5\% & AQA  \\
Clotho-AQA \cite{lipping2022clotho} & 7.4 & 2.5\% & AQA \\
CochlScene \cite{jeong2022cochlscene} & 169 & 2.5\% & ASC \\
\bottomrule
\end{tabular}
\end{table}

Table~\ref{table:trainsets} lists the training datasets covering ASR, AQA, and ASC tasks.
The audio from the LibriSpeech dataset was augmented using the speed perturbation technique~\cite{ko2015audio}.
LibriSpeech signals were speed-perturbed with 40\% probability, using a factor of 1.1 for 20\% of the samples and 0.9 for another 20\%.
For computational efficiency, utterances longer than 25 seconds after speed perturbation were excluded from training.
For the Clotho-AQA dataset, which contains answers from three annotators, we retained only instances for which at least two annotators agreed and used the majority label as the target.
The training configuration was determined through preliminary experiments, using the LibriSpeech dev-other word error rate (WER) as the primary metric.

All models were trained using AdamW for 80,000 steps with a cosine learning-rate scheduler.
The peak and minimum learning rates were set to $5 \times 10^{-4}$ and $5 \times 10^{-6}$, respectively, and the number of warm-up steps was set to 2,000.
The batch size was set to 64.

\subsection{Experimental Results and Discussion}

Table~\ref{table:mler} reports WERs for LibriSpeech and classification error rates for Clotho-AQA and CochlScene.
In the table, ``Encoder-only'' and ``Symbiotic'' denote architectures that keep the backbone LLM frozen.
Compared with ``Encoder-only'', the proposed ``Symbiotic'' substantially reduced error rates, especially on non-ASR tasks.
The high error rates on non-ASR tasks for ``Encoder-only'' suggest that adapting only the input embeddings provides limited control over the frozen LLM.

To assess how well ``Symbiotic'' controls the backbone LLM without fine-tuning, we compare it with ``Monolithic'', which adapts the backbone LLM directly.
We observed that ``Symbiotic'' closely approached this reference while reducing the number of parameters activated during audio prefilling from 1,108M to 552M.
The remaining gap on CochlScene may stem from our CNN-centric injector design; exploring alternative designs is promising future work.

Prefilling speed was measured on the LibriSpeech ``dev-other'' dataset using an H100 accelerator with a batch size of 4.
The proposed approach processed all utterances in 157.1 seconds, whereas the monolithic architecture required 199.8 seconds.
Although our injector performs subsampling at a later stage in the pipeline and therefore requires more FLOPs in theory, it still achieved a shorter wall-clock time.
We attribute this improvement to fewer kernel invocations and a cache-friendly parameter size.
In real-world applications with longer system prompts, the additional FLOPs may be amortized because the monolithic architecture relies on quadratic-time self-attention modules for prefilling.

\begin{table}[!t]
\vspace{-6pt}
\caption{Benchmark scores on text tasks.}
\label{table:text_task_results}
\centering
\begin{tabular}{lrr}
\toprule
{\bf Task} & 
Frozen LLM & 
Audio FT\\
\midrule
WikiText-2 (BPB; $\downarrow$) &
{\bf 0.88} &
1.32 \\
HellaSwag (Accuracy; $\uparrow$) &
{\bf 47.3} &
38.2 \\
GSM8K (Accuracy; $\uparrow$) &
{\bf 57.8} &
2.0 \\
\bottomrule
\end{tabular}
\end{table}

To measure the impact of catastrophic forgetting, we evaluated both the original and the fine-tuned (``Monolithic'') models on the text tasks.
WikiText-2 \cite{merity2016pointer} and HellaSwag \cite{zellers2019hellaswag} assess next-token prediction, whereas GSM8K \cite{cobbe2021gsm8k} assesses reasoning ability.
Table~\ref{table:text_task_results} shows that catastrophic forgetting due to audio-instruction fine-tuning (``Audio FT'') was severe.
By contrast, because the symbiotic architecture keeps the backbone LLM frozen, its text-only task scores are identical to that of ``Frozen LLM'' by construction.
Unlike multi-dataset training, which may compromise target-task performance~\cite{huang2022modality}, our architecture preserves the original text-only performance without sacrificing audio-task quality.

Finally, Table~\ref{table:ablation} reports LibriSpeech WERs as we incrementally add the two stabilization techniques, noisy RoPE and KV scale matching.
The last column reports the WER on a long-utterance set comprising 86 utterances longer than 25 seconds from the LibriSpeech development and test partitions; utterances of these lengths were excluded from training.

Noisy RoPE substantially improved performance, reducing the WER on long utterances from 120.68\% to 10.49\%.
KV scale matching further reduced it to 4.70\% while also improving the results on the standard splits.
These results demonstrate that both techniques improve robustness to sequence lengths unseen during training.

\begin{table}[!t]
\vspace{-6pt}
\caption{Effect of training techniques on ASR WER (\%). The four LibriSpeech values correspond to dev-clean/dev-other/test-clean/test-other.}
\label{table:ablation}
\centering
\begin{tabular}{lrr}
\toprule
{\bf Model} & 
{\bf LibriSpeech} & 
{\bf Long}\\
\midrule 
Architecture-only &
5.68/7.44/6.97/6.37 &
120.68 \\
+ Noisy RoPE &
2.13/3.97/2.21/4.09 & 10.49 \\
+ KV Scale Matching &
1.91/3.73/2.07/3.91 & 4.70 \\
\bottomrule
\end{tabular}
\end{table}

\section{Conclusions}

In this paper, we proposed a symbiotic architecture that couples an audio encoder with a backbone LLM through the LLM's short-term memory, specifically its KV cache.
By construction, it decouples audio-injection cost from the backbone width and eliminates catastrophic forgetting, since the backbone LLM is never updated.
Our experiments showed that, in the compact regime, the proposed architecture approaches a fine-tuned reference on audio tasks while activating fewer parameters during audio prefilling and preserving the backbone LLM's text-only performance exactly.
Among methods that keep the backbone frozen, it substantially outperforms the encoder-only SLM baseline, especially on non-speech audio tasks.

Since we focused on compact ALMs, our experiments used a single small backbone; verifying the scalability advantage with larger backbones remains important.

\FloatBarrier
\clearpage

\bibliographystyle{IEEEbib}
\bibliography{strings,refs}

\end{document}